\documentclass[reprint,
superscriptaddress,
 amsmath,amssymb,
 aps,
]{revtex4-1}

\usepackage{color}
\usepackage{graphicx}% Include figure files
\usepackage{dcolumn}% Align table columns on decimal point
\usepackage{bm}% bold math
\usepackage{graphicx}
\usepackage{epstopdf}
\usepackage{gensymb}
\usepackage{float}

\begin{document}

\title{Semi adsorption-controlled growth window for half Heusler FeVSb epitaxial films}

\author{Estiaque H. Shourov}
\affiliation{Materials Science and Engineering, University of Wisconsin--Madison}

\author{Ryan Jacobs}
\affiliation{Materials Science and Engineering, University of Wisconsin--Madison}

\author{Wyatt A. Behn}
\affiliation{Physics, University of Wisconsin--Madison}

\author{Zachary J. Krebs}
\affiliation{Physics, University of Wisconsin--Madison}

\author{Chenyu Zhang}
\affiliation{Materials Science and Engineering, University of Wisconsin--Madison}

\author{Patrick J. Strohbeen}
\affiliation{Materials Science and Engineering, University of Wisconsin--Madison}

\author{Dongxue Du}
\affiliation{Materials Science and Engineering, University of Wisconsin--Madison}

\author{Paul M. Voyles}
\affiliation{Materials Science and Engineering, University of Wisconsin--Madison}

\author{Victor W. Brar}
\affiliation{Physics, University of Wisconsin--Madison}

\author{Dane D. Morgan}
\affiliation{Materials Science and Engineering, University of Wisconsin--Madison}

\author{Jason K. Kawasaki}
\email{jkawasaki@wisc.edu}
\affiliation{Materials Science and Engineering, University of Wisconsin--Madison}

\date{\today}
\begin{abstract}

The  electronic, magnetic, thermoelectric, and topological properties of Heusler compounds (composition $XYZ$ or $X_2 YZ$) are highly sensitive to stoichiometry and defects. Here we establish the existence and experimentally map the bounds of a \textit{semi} adsorption-controlled growth window for semiconducting half Heusler FeVSb films, grown by molecular beam epitaxy (MBE). We show that due to the high volatility of Sb, the Sb stoichiometry is self-limiting for a finite range of growth temperatures and Sb fluxes, similar to the growth of III-V semiconductors such as GaSb and GaAs. Films grown within this window are nearly structurally indistinguishable by X-ray diffraction (XRD) and reflection high energy electron diffraction (RHEED). The highest electron mobility and lowest background carrier density are obtained towards the Sb-rich bound of the window, suggesting that Sb-vacancies may be a common defect. Similar \textit{semi} adsorption-controlled bounds are expected for other ternary intermetallics that contain a volatile species $Z=$\{Sb, As, Bi\}, e.g., CoTiSb, LuPtSb, GdPtBi, and NiMnSb. However, outstanding challenges remain in controlling the remaining Fe/V ($X/Y$) transition metal stoichiometry.

\end{abstract}

%\pacs{Valid PACS appear here}

\maketitle

The remarkable success of III-V compound semiconductor epitaxial films is due in large part to the existence of a thermodynamically adsorption-controlled growth window, in which the stoichiometry is self-limiting \cite{stormer1981influence, tsui1982twodimensional, tsao2012materials, cho1975molecular, pfeiffer1989electron}. For these materials, due to the high volatility of the group $V$ = \{As, Sb, N, or P\} species, there exists a finite range of temperatures and fluxes, called the ``growth window,'' in which only the stoichiometric composition of group $V$ incorporates into the film, while the excess group $V$ species escapes into the vapor. This remarkable level of stoichiometry control, precise to near parts per billion, has enabled the growth of semiconductors with record high electron mobility \cite{pfeiffer1989electron, dingle1978electron}, fundamental physical discoveries such as the integer and fractional quantum Hall effects \cite{tsui1982twodimensional, klitzing1980new}, ultrafast transistors \cite{awano1989electron}, and optoelectronics. Similar adsorption-controlled growth windows have been identified for binary chalcogenides (CdTe, SnSe, Bi$_2$Se$_3$, WTe$_2$, volatile chalcogen), nitrides (NbN, Ta$_2$N, volatile nitrogen), oxides (ZnO, TiO$_2$, volatile oxygen), and in select cases, ternary transition metal oxides using a volatile binary oxide or metalorganic precursor (BiFeO$_3$ \cite{ihlefeld2007adsorption}, BaSnO$_3$ \cite{paik2017adsorption, prakash2017adsorption}, SrTiO$_3$ \cite{jalan2009molecular}, SrVO$_3$ \cite{brahlek2015accessing}). 

Heusler compounds are another important class of materials, of great interest for their magnetic, thermoelectric, and topological properties \cite{graf2011simple, palmstrom2016heusler}. Heuslers are ternary intermetallics with composition $XYZ$ (half Heusler) or $X_2 YZ$ (full Heusler), where $X$ and $Y$ are transition or rare earth metals and $Z$ is typically a main group metal. Their electronic and magnetic properties are highly sensitive to nonstoichiometry and the associated defects \cite{yonggang2017natural, picozzi2007polarization, picozzi2004role, ougut1995band, larson2000structural}. However, it remains an outstanding challenge to control the stoichiometry to ``electronic-grade'' quality. For example, while the intrinsic carrier concentration of silicon is $n_i \sim 10^{10}$ cm$^{-3}$ at room temperature, typical experimental carrier concentrations for semiconducting half Heuslers are typically $\sim 10^{19}$ to $10^{21}$ cm$^{-3}$ due to defects and nonstoichiometry, which are difficult to control to better than $1\%$ \cite{kawasaki2019heusler}. In select cases it has been shown that several Sb-containing Heuslers -- including CoTiSb \cite{kawasaki2018simple, kawasaki2014growth}, NiMnSb \cite{bach2003molecular, turban2002growth}, LuPtSb \cite{patel2014surface}, LaPtSb \cite{du2019high}, and LaAuSb \cite{strohbeen2019electronically} -- can be grown with an excess Sb flux, in which the ratio of Sb to $(X+Y)$ is self-limiting. Since the stoichiometry of one out of three elements is self-limiting, this can be called \textit{semi} adsorption control. However, the thermodynamic basis and the bounds of the growth window for Heuslers have not yet been established.

Here we establish the thermodynamic basis and experimentally map the bounds of the \textit{semi} adsorption-controlled growth window for FeVSb films, grown by molecular beam epitaxy (MBE). FeVSb is a semiconducting half Heusler compound with a large thermoelectric power factor and is the parent compound for a number of doped and nanostructured high efficiency thermoelectrics \cite{young2000thermoelectric}. We show that within a finite range of temperatures and Sb fluxes, the Sb stoichiometry is self-limiting and the resultant single-crystalline FeVSb films are nearly structurally indistinguishable, as revealed by reflection high energy electron diffraction (RHEED) and X-ray diffraction (XRD). Hall effect measurements reveal that the electron mobility is optimized near the Sb-rich bound of the window, suggesting that Sb-vacancies may be a common defect. However, outstanding challenges remain for controlling the Fe/V $(X/Y)$ stoichiometry, which is not self-limiting for growth using elemental transition metal fluxes.

\begin{figure*}
  \centering
    \includegraphics[width=1\textwidth]{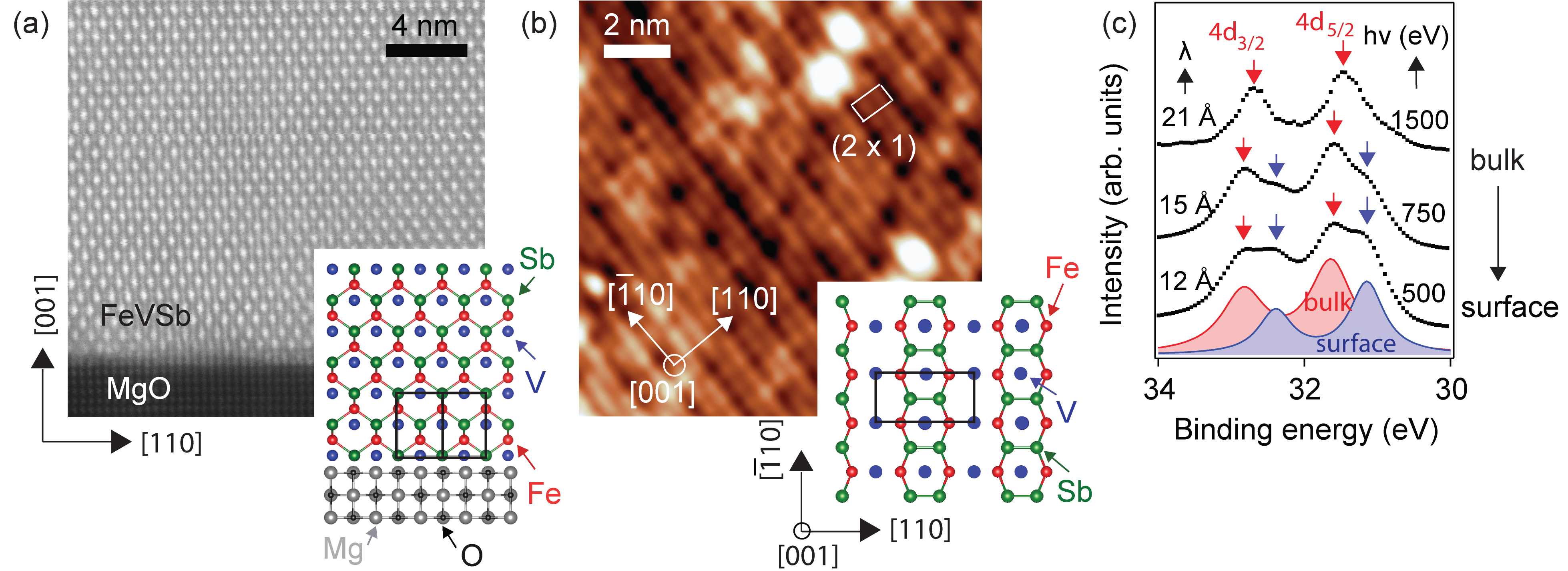}
    \caption{\textbf{Bulk and surface structure of FeVSb (001) films.} (a) Cross sectional HAADF-STEM image of the FeVSb/MgO interface. Crystallographic directions are referenced to the FeVSb unit cell. Inset: Structural model of FeVSb/MgO (001). (b) Empty states STM image (500 mV sample bias, 3 nA tunnel current) of the FeVSb surface showing a $(2 \times 1)$ surface reconstruction. These samples were grown at a temperature of $\approx$ 450$\degree$C and Sb/V flux ratio of 2.2. Inset: Model of the surface reconstruction, characterized by Sb-Sb dimerization. (c) Sb 4d core level evolution as a function of photon energy, showing evidence for surface Sb-Sb dimerization. The estimated photoelectron mean free path $\lambda$ is derived from the universal curve \cite{seah1979quantitative}. Shaded curves show a Voigt fit to the $h\nu=500$ eV data.}
    \label{surface}
\end{figure*}

FeVSb films were grown in a custom MBE system on MgO (001) substrates by co-deposition from elemental effusion cells. The lattice mismatch between FeVSb and MgO is 2.19$\%$ tensile when rotated 45$\degree$ around the c--axis. MgO substrates (MTI Corporation) were annealed at 700$\degree$C in ultrahigh vacuum (base pressure less than $2\times10^{-10}$ Torr) until the appearance of a bright $(1\times 1)$ reflection high energy electron diffraction (RHEED) pattern, after which the temperature was increased to the desired growth temperature. The substrate temperature was measured using a thermocouple and calibrated for each sample puck to the oxide desorption temperature (500$\degree$C) and melting point (712$\degree$C) of GaSb. Fe and V fluxes of $7.9 \times 10^{12}$ atoms/(cm$^{2}$s) were supplied from a standard and a high temperature cell, respectively. The Sb flux was supplied from a thermal cracker cell with the cracker region operating at 1200$\degree$C, corresponding to a mixed flux of molecular Sb$_2$ and atomic Sb$_1$. All fluxes were measured by an \textit{in situ} quartz crystal microbalance (QCM) that was calibrated to each cell geometry by \textit{ex situ} Rutherford backscattering spectroscopy (RBS). 

An overview of the bulk and surface structure of our FeVSb/MgO (001) films is shown in Fig. \ref{surface}. The Z-contrast high angle annular dark field scanning transmission electron microscopy (HAADF-STEM) image in Fig. \ref{surface} (a) confirms an epitaxial FeVSb film on MgO with a 45 degree rotated cube-on-cube relationship, i.e. FeVSb $(001)[110] \parallel$ MgO $(001)[100]$. Empty states scanning tunneling microscopy (STM, Fig. \ref{surface} (b)) images reveal a $(2\times 1)$ surface reconstruction, similar to what has been observed for other half Heusler (001) surfaces \cite{kawasaki2018simple, patel2016surface, bach2003molecular, turban2002growth}. To understand the origin of this $(2\times 1)$ surface reconstruction, we performed photon energy-dependent photoemission spectroscopy measurements at beamline 29-ID of the Advanced Photon Source (Fig. \ref{surface} (c). We find that with decreasing photon energy (increasing surface sensitivity), the Sb $4d$ core level shows a secondary component with decreased binding energy that is localized to the surface. We attribute this secondary component to surface Sb-Sb dimerization \cite{kawasaki2018simple}. A proposed model of the surface atomic structure is shown in Fig. \ref{surface}(b inset), characterized by Sb-Sb dimerization. Some concentration of surface V vacancies is expected based on surface charge neutrality \cite{kawasaki2018simple}; however, quantifying this effect is beyond the scope of the current study. Such vacancies are localized the surface and are expected to have negligible affect on the bulk properties. Further TEM, STM, and photoemission measurement details are found in the Supplemental Information.

\begin{figure}
    \centering
    \includegraphics[width=0.45\textwidth]{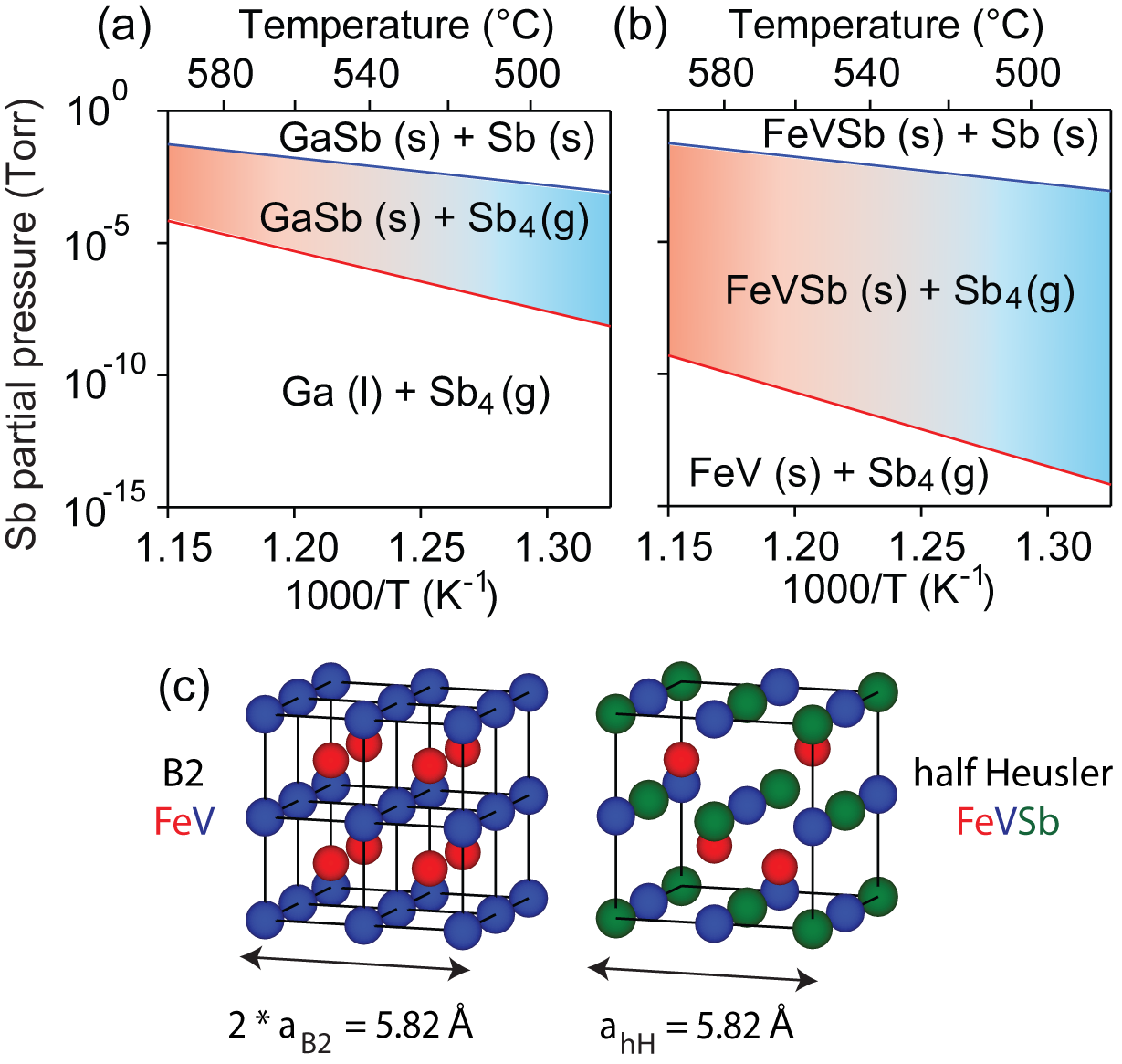}
    \caption{\textbf{Thermodynamics of FeVSb and GaSb adsorption-controlled growth.} (a-b) Ellingham diagrams for GaSb and FeVSb. In both plots, the upper curves (blue) represent the change in Gibbs free energy for antimony sublimation Sb(s) $\Leftrightarrow$ $\frac{1}{4}$Sb$_{4}$(g). The lower curves (red) are the change in free energies for decomposition of solid GaSb or FeVSb, respectively. The shaded regions bounded by Sb sublimation and FeVSb (GaSb) decomposition define the growth window, in which solid FeVSb or GaSb are in equilibrium with antimony vapor, hence the Sb stoichiometry of the solid is self-limited. See text for further details. (c) Crystal structures for FeV and FeVSb, with an expected $2a_{B2} \approx a_{hH}$ epitaxial relationship.}
    \label{thermo}
\end{figure}

We first establish the thermodynamic basis for \textit{semi} adsorption-controlled growth of FeVSb, in which the Sb stoichiometry is self-limiting. Fig. \ref{thermo} compares the Ellingham diagram for FeVSb with that of GaSb, a III-V semiconductor that is routinely grown by adsorption-control. The common upper bound (blue curves), represents the change in Gibbs free energy for sublimation of antimony Sb(s) $\Leftrightarrow$ $\frac{1}{4}$Sb$_{4}$(g), as obtained from tabulated thermodynamic data \cite{barin1995thermochemical}. For temperature and Sb partial pressure combinations above this curve, solid antimony is expected to precipitate on the surface of the film. Below this curve, excess antimony is not expected to incorporate into the film, and instead escape into the vapor. The lower curves (red) represent the decomposition reactions GaSb(s) $\Leftrightarrow$ Ga(l) + $\frac{1}{4}$Sb$_{4}$(g) [Fig. \ref{thermo}(a)] and FeVSb(s) $\Leftrightarrow$ FeV(s) + $\frac{1}{4}$Sb$_{4}$(g) [Fig. \ref{thermo}(b)], respectively. The GaSb curve is obtained from completely from tabulated thermodynamic data \cite{barin1995thermochemical}. The FeVSb curve is calculated by combining density functional theory (DFT) calculations for FeVSb and FeV with tabulated thermodynamic data for Sb sublimation (see Supplemental Information). For temperature and Sb partial pressure combinations below these curves, FeVSb and GaSb are expected to decompose into their binary and elemental constituents, respectively. The regions bounded by the FeVSb (GaSb) decomposition and Sb sublimation curves define the expected growth windows for FeVSb and GaSb, respectively. Here, solid FeVSb or GaSb form with the stoichiometric composition of Sb, while the excess Sb escapes into the vapor. Based on these thermodynamic calculations, we expect that compared to GaSb, FeVSb should be stable at higher growth temperatures and lower Sb partial pressures. Few percent changes in the Fe/V stoichiometry produce minimal changes to the expected Ellingham diagram (Supplemental Fig. 1).

\begin{figure*}
    \centering
    \includegraphics[width=.9\textwidth]{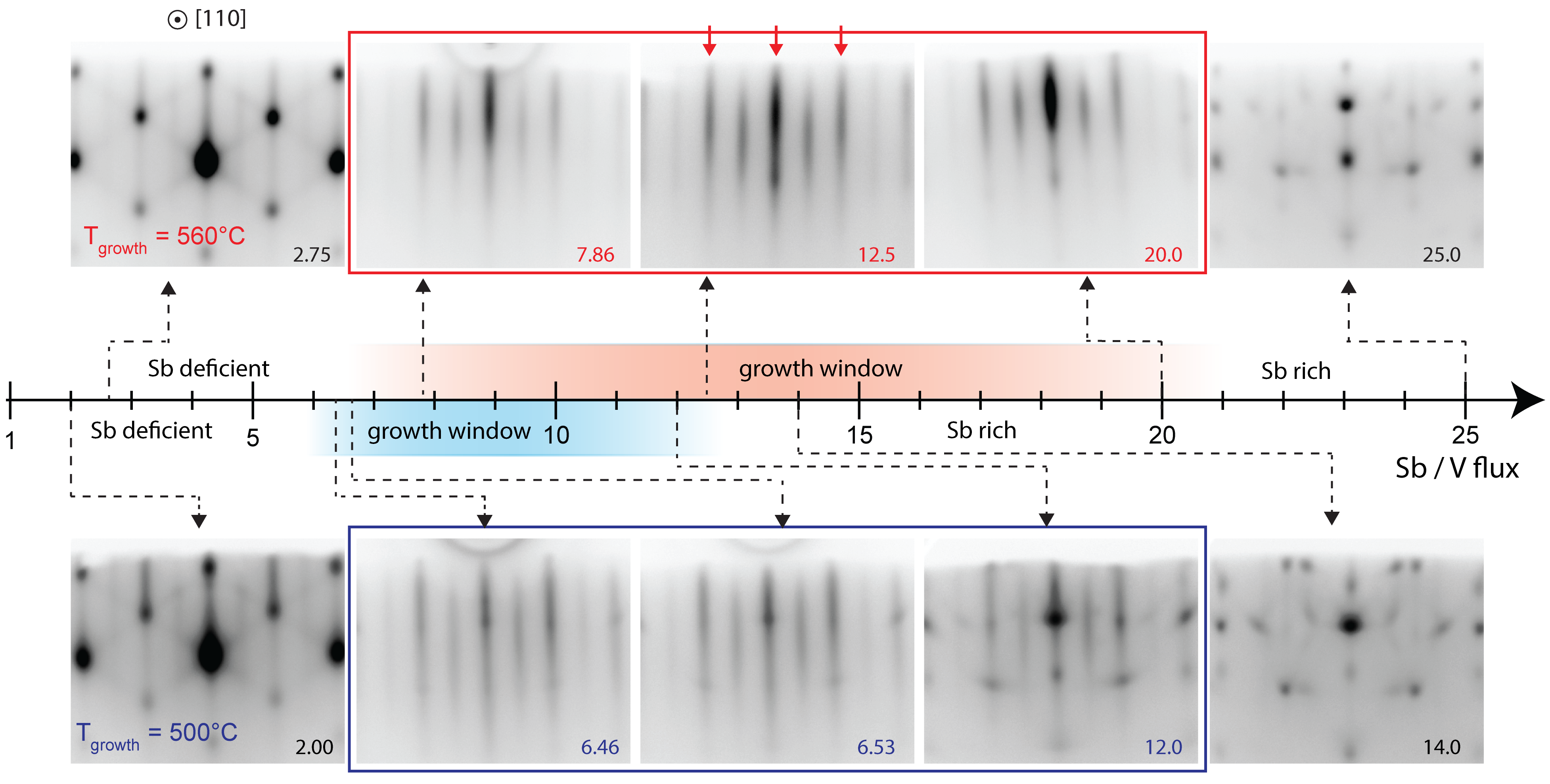}
    \caption{RHEED patterns for beam oriented along the [110] azimuth of FeVSb films grown at 500$\degree$C (bottom) and at 560$\degree$C (top) with varying Sb/V flux. Red arrows mark the bulk 1$\times$ reflections. The Sb deficient region corresponds to mixed phases of FeVSb (half-heusler) and FeV (B2), which are epitaxial with one another. The Sb rich region corresponds to excess Sb islands formed on the FeVSb surface. Within the growth window, the RHEED patterns exhibit a streaky (2$\times$1) pattern characteristic of half Heusler surfaces.} 
    \label{rheed}
\end{figure*}

We now experimentally map the bounds of the FeVSb \textit{semi} adsorption-controlled window. Fig. \ref{rheed} (bottom) shows the RHEED patterns for samples grown at fixed temperature of 500$\degree$C, as a function of Sb/V atomic flux ratio. For Sb/V flux ratio less than 5, the RHEED patterns are spotty, indicative of three-dimensional island formation. Based on the ternary Fe-V-Sb phase diagram, we expect highly Sb-deficient conditions to correspond to an epitaxial coexistence of FeV (B2 structure, $a_{B2} = 2.91$ \AA) and FeVSb (half Heusler structure, $a_{hH} = 5.82$ \AA), where $a_{hH} \approx 2 a_{B2}$ \cite{romaka2012interaction} (Fig. \ref{thermo}(c)). There exists a range of moderate Sb/V fluxes, from approximately 5 to 12, in which a streaky $(2\times 1)$ RHEED pattern is observed, indicating smooth crystalline surfaces with FeVSb in half Heusler structure. This range of moderate Sb/V fluxes defines the growth window. Above an Sb/V flux ratio of 12, we observe additional spots in the RHEED pattern, indicative of Sb islands forming on the surface. Similar trends are observed for growth at a higher temperature of 560$\degree$C, in which the bounds of the window are shifted towards higher Sb/V flux ratio as shown in Fig. \ref{rheed} (top).

\begin{figure} 
    \centering
    \includegraphics[width=0.4\textwidth]{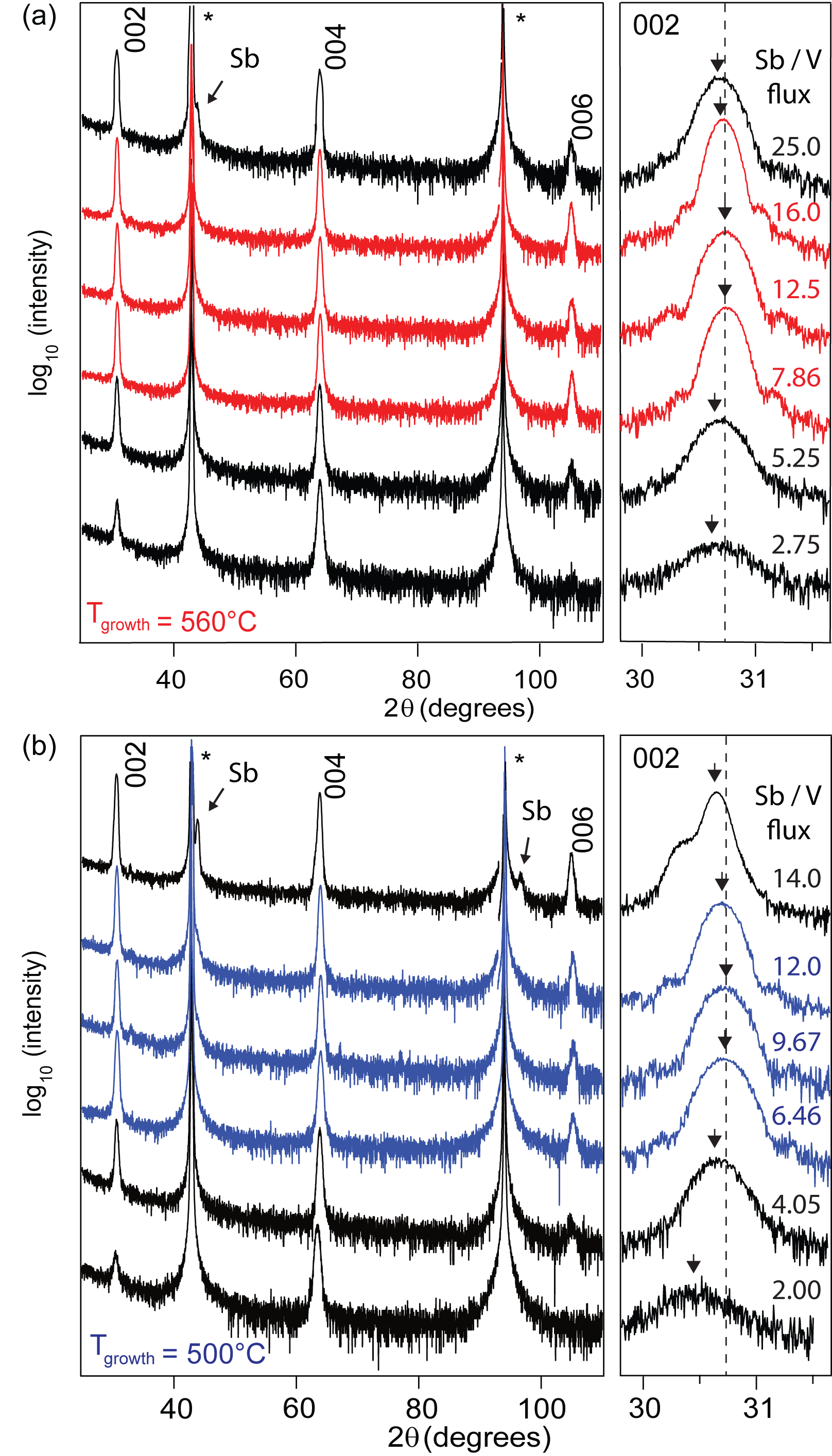}
    \caption{\textbf{Structural growth window.} Left panels: $\theta-2\theta$ X-ray diffraction scans (Cu $K\alpha$) for samples grown at (a) 560$\degree$C and (b) 500$\degree$C as a function of Sb/V atomic flux ratio. Substrate reflections are marked by asterisks. $00l$-type reflections are observed corresponding to an epitaxial FeVSb film. For very Sb-rich conditions, reflections from a precipitated Sb phase are observed at 2$\theta$ = 43.93$\degree$ and 96.83$\degree$. Right panels: higher resolution scans of the FeVSb $002$ reflection, tracking changes in the out-of-plane lattice parameter.}
    \label{structure}
\end{figure}

We further quantify the growth window for half Heusler phase by bulk-sensitive X-ray diffraction. Fig. \ref{structure} shows $\theta-2\theta$ scans (Cu $K\alpha$) for samples grown at 500$\degree$C and at 560$\degree$C. In all samples we observe only $00l$-type FeVSb reflections, corresponding to epitaxial FeVSb films. We focus on the samples grown at 500 $\degree$C; samples grown at 560$\degree$C show similar qualitative behavior. Starting from an Sb/V atomic flux ratio of 2.0, with increasing relative Sb flux the relative intensity of the $002$ and $006$ reflections increases, and there is a shift in the $002$ reflection towards higher angle, corresponding to a decrease in the out-of-plane lattice parameter. In this region, we attribute the changes in structure factor and lattice parameter to two possible microstructures: (1) a mixture of FeV (B2) and FeVSb (half Heusler) phases under very Sb-deficient conditions, and (2) Sb vacancies near stoichiometric conditions. Firstly, for extremely low Sb flux conditions, a mixture of FeV (cubic B2) and FeVSb (half Heusler) phases is expected from the ternary phase diagram, in which the two phases are epitaxial to one another (a$_{hH}$ $\approx 2 a_{B2}$) \cite{romaka2012interaction}. The B2 phase has a much smaller 001$_{B2}$/002$_{B2}$ structure factor ratio than the equivalent half Heusler 002$_{hH}$/004$_{hH}$. With increasing Sb flux, the relative volume fraction of half Heusler FeVSb to B2 FeV increases, consistent with the observed increase in the 002/004 intensity ratio (Supplemental Fig. 3). Secondly, for Sb fluxes near stoichiometry, Sb vacancies may be present. Our structure factor calculations show that Sb vacancies in the half Heusler structure decrease the 002/004 and 006/004 structure factor ratio (Supplemental), consistent with the measured trends.

\begin{figure}
    \centering
    \includegraphics[width=.48\textwidth]{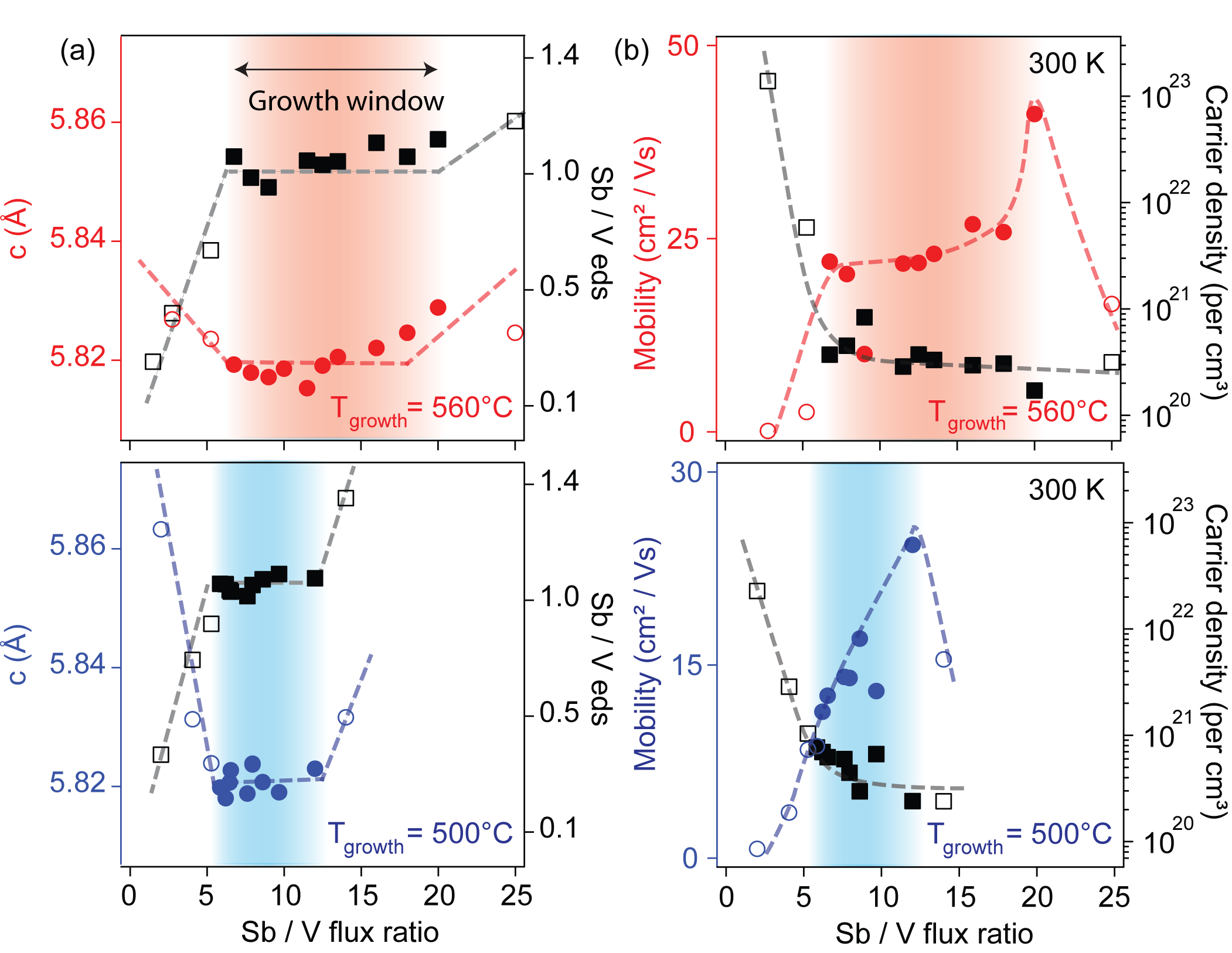}
    \caption{\textbf{Structural and electrical transport window.} (a) Out of plane lattice parameter (circles, left axis) and Sb/V composition (squares, right axis) as a function of Sb/V atomic flux ratio, at growth temperatures of 560$\degree$C and 500$\degree$C. The out-of-plane lattice parameter was extracted from the peak position of the 002 reflection in X-ray diffraction. The Sb/V ratio was determined by energy dispersive X-ray spectroscopy measurements, which are calibrated to a known standard by Rutherford Backscattering Spectrometry (RBS). Within the structural growth window, both the lattice constant and Sb/V stoichiometry are constant within error. (b) Electron mobility (left axis) and density (right axis) at 300K extracted from the Hall effect at field greater than 2T.}
    \label{transport}
\end{figure}

Over a wide range of intermediate Sb/V flux ratios, from approximately 6 to 12, both the relative intensities of the $002$ and $006$ reflections and the peak position of the $002$ reflection are constant. This broad range of of Sb/V defines the growth window, and the extracted lattice parameters from the $002$ and $004$ reflections are shown in Fig. \ref{transport}(a). Here the lattice parameter for films within the window plateaus to a value of $\approx$5.82 {\AA}, in good agreement with previously reported lattice constant of 5.826 {\AA} for bulk samples \cite{jodin2004effect}. We attribute variations in the lattice parameter to variations in Fe/V composition, which in the present study vary by no more than 5 percent. Samples within this window show Kiessig fringes around the $002$ reflections (Fig. \ref{structure}), indicative of smooth interfaces. Within this window, the Sb stoichiometry plateaus to a constant Sb/V=1.0, as measured by energy dispersive x-ray spectroscopy (EDS) that we calibrate in absolute scale to Rutherford Backscattering Spectrometry (RBS) measurements on a few select samples (Fig. \ref{transport}(a), see Supplemental Information for measurement details and analysis methods).

For Sb/V flux greater than 12, we observe precipitation of a secondary phase in XRD (Fig. \ref{structure}) and an increase in the out-of-plane lattice parameter (Fig. \ref{transport}(a)). This defines the Sb-rich bound of the window. We find that the Sb-rich bound observed by XRD (Sb/V $\sim 12-14$, Fig. \ref{structure}) is higher than the bound defined by the onset of spottiness in the RHEED pattern (Sb/V $\sim 10-12$, Fig. \ref{rheed}). We attribute this discrepancy to Sb precipitates localized to the surface, which can be detected by surface-sensitive RHEED but are not detected by bulk-sensitive X-ray diffraction.

Fig. \ref{transport}(b) shows the room temperature Hall electron mobility and density for samples grown at 500 and 560$\degree$C. These measurements were performed using a Quantum Design Physical Properties Measurement System (PPMS) in a Van der Pauw geometry using annealed indium contacts. At low fields ($\mu_{0}$H $<$ 0.7 T) the Hall resistance (R$_{xy}$) for some samples showed slight nonlinearities (Supplemental Fig. 5), which we attribute to ferromagnetic impurities that arise from slight Fe-nonstoichiometry. We extract the carrier density by fitting a single band model to the high field ($>$ 2 T) regime, where all samples showed linear R$_{xy}$ vs $\mu_{0}$H. Starting from Sb-deficient conditions, we find that a sharp increase in mobility and decrease in carrier density as the Sb flux is increased into the growth window. Within the growth window, the carrier mobility increases with Sb/V flux and peaks at the Sb-rich bound of the structural growth window. This behavior suggests that Sb vacancies may be a low formation energy defect. While Sb vacancies are not readily detected by diffraction or composition measurements, they are expected to strongly contribute to carrier scattering.

Films grown at a higher temperature of 560$\degree$C reach a higher peak mobility than films grown at 500$\degree$C, which we attribute to increased atomic site ordering at higher growth temperature. The maximum mobility of 41 cm$^{2}$/Vs at 300 K for FeVSb films grown inside a growth window is comparable to previously reported mobilities of epitaxial half Heusler films grown on MgO (001), but is smaller than the highest mobilities reported for films grown on III-V substrates such as InP (001) \cite{patel2016surface, kawasaki2013epitaxial, kawasaki2014growth}. We attribute the reduced mobility to antiphase domains that form as a result of growing a (001)-oriented half Heusler film (2-fold rotational symmetry) on a rocksalt MgO (001) substrate (4-fold rotation). Growth on a 2-fold surface of a (001)-oriented III-V surface is expected to yield higher mobility films \cite{kawasaki2014growth}.

Supplemental Fig. 6 shows the temperature dependence of the resistivity, mobility and carrier density of two 560$\degree$C films, one grown in the middle (Sb/V flux 11.5) and the other at the Sb rich bound (Sb/V flux 20.0) of the structural growth window. For both samples, the electron density and resistivity show a weak temperature dependence, and the magnitude of the electron density of $10^{20}$ cm$^{-3}$ is consistent with degenerate doping. We attribute this doping to antiphase domains induced by the 4-fold MgO (001) substrate and to a few percent deviation in the Fe/V stoichiometry, which is not self-limited. The mobilities increase with decreasing temperatures as expected with acoustic phonon freeze out at low temperatures. Further measurements are required to quantify and identify the defect-induced scattering mechanisms for films grown in the middle of the structural window versus the mobility-optimized films grown at near the Sb-rich bound of the structural window.

\begin{figure}
    \centering
    \includegraphics[width=.45\textwidth]{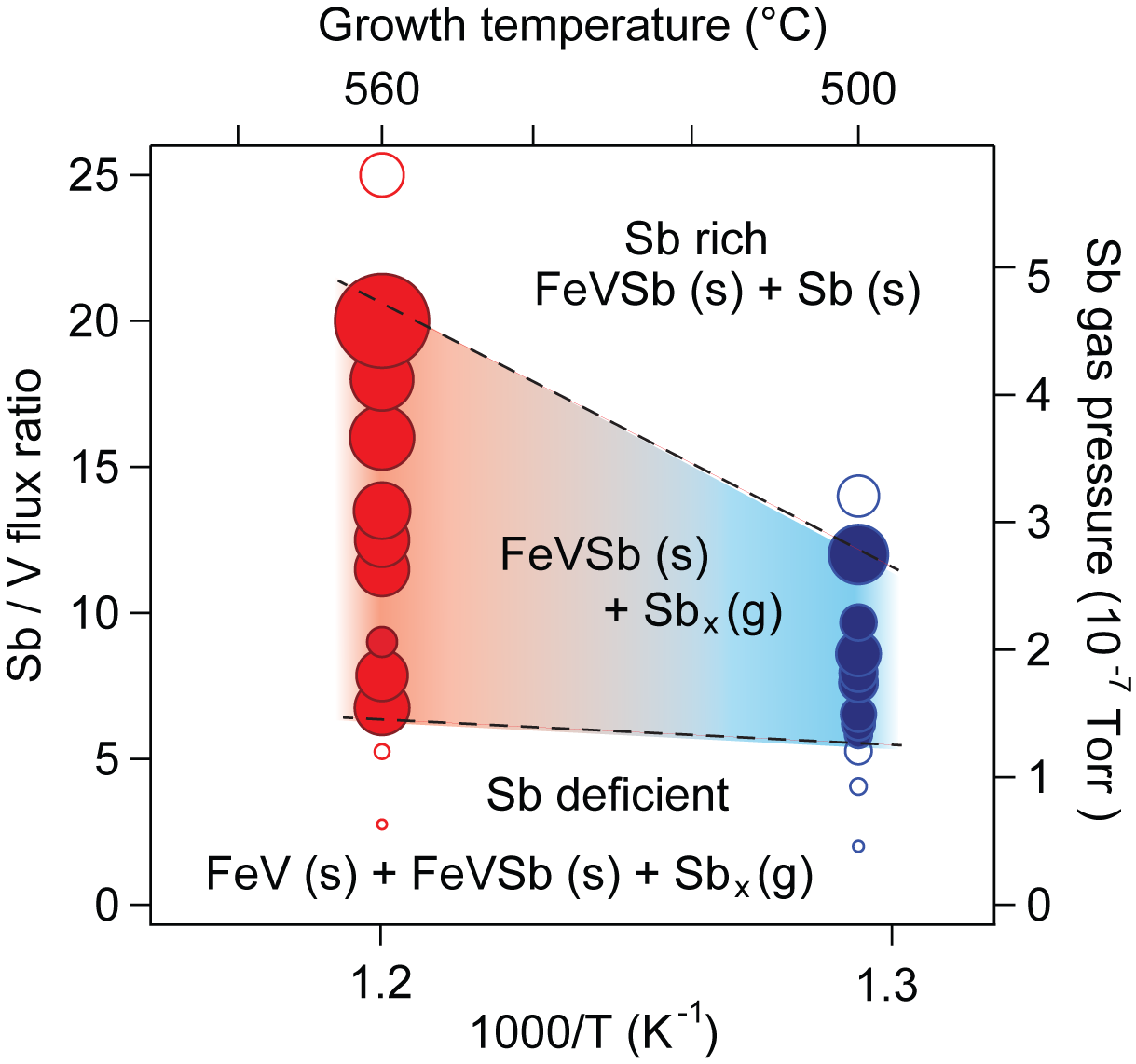}
    \caption{\textbf{Experimental phase diagram as a function of Sb/V flux ratio and inverse growth temperature.} Filled circles denote growth within the structural window as determined by RHEED, XRD, and EDS. Unfilled circles are outside the growth window. The size of the circles scales with the magnitude of the majority carrier mobility at 300K. Dotted lines are guide to the eye.}
    \label{window}
\end{figure}

Our experimental phase diagram for semi adsorption-controlled growth is summarized in Fig. \ref{window}. Here the open circles represent films outside the growth window, filled circles represent films inside the growth window, and the size of the circles scales with the magnitude of the electron mobility. To facilitate comparisons with our Ellingham diagram prediction (Fig. \ref{thermo}), we estimate the Sb partial pressure using the kinetic theory of gases \cite{farrow1995molecular}, where $p \approx \phi \sqrt{(\pi m k_{B} T)/8}$ (Fig. \ref{window} right axis). Here $\phi$ is the Sb atomic flux, $k_{B}$ is the Boltzmann constant, $T=1200^{\degree}$C is the temperature of the Sb vapor, and $m=121.8$ amu is the mass of the Sb vapor species. The Sb vapor is assumed to be atomic Sb$_{1}$ for approximation purposes, the true flux is a mixture of Sb$_{1}$ and Sb$_{2}$. We find that our experimental window is much narrower than the thermodynamic prediction: whereas the experimental window is centered around a Sb partial pressure of order $10^{-7}$ Torr and spans a factor of 2.5 to 5 (Fig. \ref{transport}(c)), the thermodynamic prediction spans several decades (Fig. \ref{thermo}(b)). We attribute this discrepancy to growth kinetics, which are not captured in the Ellingham diagram. Similar discrepancies are observed for other adsorption-controlled systems, such as GaSb. For GaSb the experimental Sb/Ga atomic flux window at $500^{\degree}$C spans approximately a factor of ten \cite{turner1993molecular, ivanov1993molecular, yano1978molecular}, much smaller than the several decade wide prediction from the Ellingham diagram (Fig. \ref{thermo}(a)). A more complete view of the growth window requires the includion of kinetics, which has recently been applied to the MBE growth of several transition metal oxides \cite{smith2017exploiting}.

In summary, we have established the thermodynamic basis and experimentally mapped the semi adsorption-controlled growth window for half Heusler FeVSb films, in which the Sb stoichiometry is self-limiting. Similar Sb adsorption-controlled windows are expected for CoTiSb, LuPtSb, and NiMnSb, which have also been grown previously by MBE with excess Sb fluxes \cite{logan2016observation, kawasaki2018simple, bach2003molecular}, but whose growth window bounds have yet to be quantified. It remains an outstanding challenge to control the Fe/V stoichiometry, which is not self-limited when using atomic Fe and V fluxes from effusion cells. Recent demonstrations MBE-grown LiZnSb, in which all three atomic species are volatile, suggest that it may be possible to control the full stoichiometry of a ternary Heusler compound \cite{du2020control}. However, for transition metal based Heuslers such as FeVSb, control of the $X/Y$ transition metal ratio may require replacing one or both of the elemental transition metal sources with a volatile metalorganic precursor.

\section{Acknowledgments}

We thank Fanny Rodolakis and Jessica L. McChesney for technical support during synchrotron photoemission measurements. This work was supported by the CAREER program of the National Science Foundation (DMR-1752797) and by the SEED program of the Wisconsin Materials Research Science and Engineering Center, an NSF funded center (DMR-1720415). Ryan Jacobs and Dane Morgan were supported by the Department of Energy (DOE) award number DE-SC0020419. We gratefully acknowledge the use of X-ray diffraction and electron microscopy facilities supported by the NSF through the University of Wisconsin Materials Research Science and Engineering Center under Grant No. DMR-1720415. This research used resources of the Advanced Photon Source, a U.S. Department of Energy (DOE) Office of Science User Facility operated for the DOE Office of Science by Argonne National Laboratory under Contract No. DE-AC02-06CH11357; additional support by National Science Foundation under Grant no. DMR-0703406. We thank Professor Song Jin for the use of PPMS facilities. We thank Mark Mangus (Eyring Materials Center, Arizona State University) and Greg Haugstad (Characterization Facility, University of Minnesota) for performing RBS measurements.

\bibliographystyle{apsrev}
\bibliography{bibliography}

\end{document}